\documentclass[letter]{ieice}
\usepackage[dvipdfmx]{graphicx,xcolor}
\usepackage[fleqn]{amsmath}
\usepackage{newtxtext}
\usepackage[varg]{newtxmath}

\field{}
\title{Developing a Multi-Platform Speech Recording System Toward Open Service of Building Large-Scale Speech Corpora}
\authorlist{%
 \authorentry{Keita Ishizuka}{n}{labelA}\MembershipNumber{}
 \authorentry{Takashi Nose}{m}{labelB}\MembershipNumber{}
}
\affiliate[labelA]{The author is with the 
Graduate School of Information Sciences, %
Tohoku University, Sendai, 980-8579 Japan.}
\affiliate[labelB]{The author is with the 
Graduate School of Engineering, %
Tohoku University, Sendai, 980-8579 Japan.}

\received{2015}{1}{1}
\revised{2015}{1}{1}

\begin{document}
\maketitle
\begin{summary}
This paper briefly reports our ongoing attempt at the development of a multi-platform browser-based speech recording system. We designed the system toward a service of providing open service of building large-scale speech corpora at a low-cost for any researchers and developers related to speech processing.
The recent increase in the use of crowdsourcing services, e.g., Amazon Mechanical Turk, enable us to reduce the cost of collecting speakers in the web, and there have been many attempts to develop the automated speech collecting platforms or application that is designed for the use the crowdsourcing. However, 
one of the major problems in the previous studies and developments for the attempts is that most of the systems are not a form of common service of speech recording and corpus building, and each corpus builder is necessary to develop the system in their own environment including a web server. For this problem, we develope a new platform where both the corpus builders and recording participants can commonly use a single system and service by creating their user accounts. A brief introduction of the system is given in this paper as the start of this challenge.

\end{summary}

\section{Introduction}

The advance of machine learning techniques and computer environments have enabled the increase of the amount and variability of speech data in the speech processing research and development. There have been many attempts at collecting large-scale speech corpora in decades. For example, TIMIT \cite{zue1990speech} and SWITCHBOARD corpus \cite{godfrey1992switchboard} are well known as American speech corpora. Similarly, JNAS \cite{itou1999jnas} and Corpus of spontaneous Janapese \cite{maekawa2003corpus} are also major Japanese large-scale speech corpora developed mainly for speech recognition of diverse speakers.
In the research area of speech synthesis, the corpus size has been also increasing because of the success of deep learning-based approaches. For this purpose, Google recently published a speech corpus named LibriTTS \cite{zen2019libritts} that is a high-quality version of the former corpus of LibriSpeech \cite{zen2019libritts}.
However, most of the well-known large-scaled corpora were constructed by large groups or big companies with huge costs, and it is still not easy for researchers and developers to build the desired corpus for speech processing at a low cost.

The use of remote systems is one of the most efficient ways to reduce the recording cost. In the early 1990s, the voice across America project \cite{wheatley1991voice} conducted automated corpus collection using PCs and telephone networks, and the SWITCHBOARD and voice across Japan corpora were constructed in the same manner.
The emergence of the Internet and mobile phones also reduced the costs of collecting speech and speakers. For example, Nokia developed the Crowd Translator \cite{ledlie2009crowd} in which mobile phones were used for low-resource languages.

In this paper, we describe our ongoing development of a multi-platform browser-based speech recording and collection system to achieve an open service of building large-scale speech corpora. One of the big problems in the previous studies and developments for web-based speech collection systems is that most of the developed systems were not a form of recording service without the users' configuration. In other words, the corpus builders should develop the system in their environment including a web server. This would be a burden to the builder who wants to concentrate on building the targe corpus itself. To reduce the burden, we provide a new approach where both the corpus builders and recording participants can commonly use a single system and service by creating their user accounts. In the following sections, we describe the details of the system and service of the corpus building system.

\begin{figure*}[t]
  \begin{center}
  \includegraphics[width=1.0\hsize]{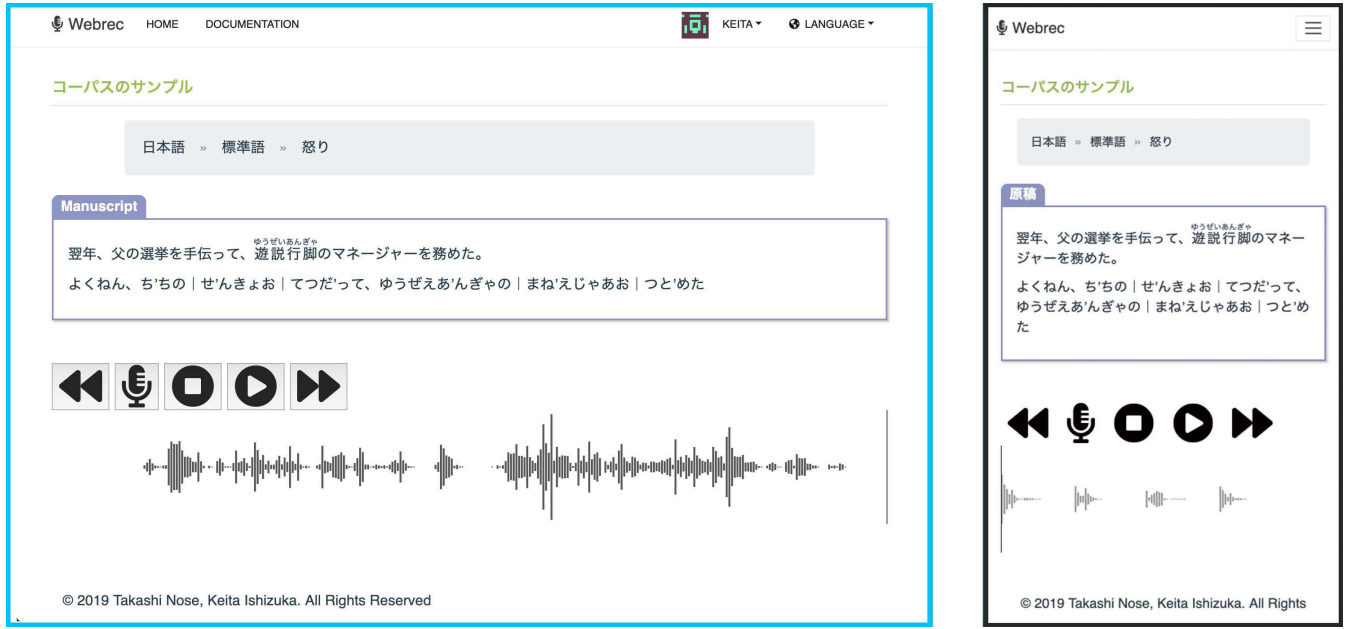}
  \end{center}
  \caption{Samples of web pages for recording Japanese speech: left for PCs and right for smartphones. Language can be switched on the configuration page (only Japanese and English in the current version).}
  \label{fig:screen}
\end{figure*}

\section{Overview of the Developed Application}

In the following sections, {\it participants} represent the persons who participate in the recording and provide their voice. {\it Builders} are the persons who use this recording system to collect speech samples and build the target corpus. The proposed system is a unified application and it provides speech recording and management by remote users, making the final corpus, checking the recording situation as well as the recording with browsers of PCs and smartphones. This section describes an overview of three functions: recording with PCs/smartphones, building a corpus, and checking the recording states.

We employed the responsible design to make the recording pages correctly render in multiple platforms, i.e., both on PCs and smartphones. The support of Web Audio API in mobile Safari and Chrome browsers enabled the recording with a smartphone. Figure \ref{fig:screen} shows the samples of recording pages in the web browsers of PCs and smartphones. The information of the corpus name, the corpus structure, the recording sentence, control buttons, and the recorded waveform are displayed in both devices.
Participants first access this page from the pages of the sentence list. After checking the recording script, they start recording by pushing the recording button and speak to the microphone. They push the stop button to finish the recording, and the recorded waveform outline is displayed below the control buttons. The speech data is sent to the server if the recording conditions are satisfied. There are currently two conditions related to speech volume and noise level. The condition for the speech volume is that the maximum and minimum values of the waveform amplitude are within the specified range. For example, the range of the absolute value of max amplitude can be set from 20,000 to 30,000 in the case of 16-bit linear quantization. The other condition is S/N which is calculated using the average powers of the beginning and whole of the waveform. When the conditions for the recorded speech is satisfied, participants can move to the next sentence.

Builders can start building the corpus by inputting the corpus title and explanation, providing the list of the recording scripts with a CSV format, and sending the data. After this corpus setting, the system creates the original page for checking the recording state. The builders can check the user names of the participants, the recording progress of each participant, and the recording scripts. Currently, the recording state can be monitored only by the original corpus builder.

\begin{figure*}[t]
  \begin{center}
  \includegraphics[width=0.8\hsize]{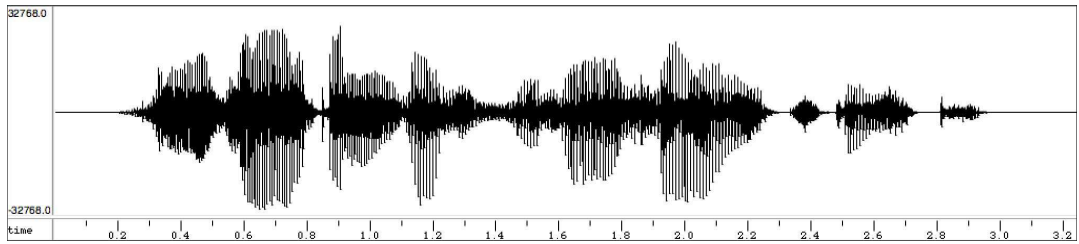}
  \end{center}
  \caption{Sample of the recorded speech waveform.}
  \label{fig:wave}
\end{figure*}

\section{Implementation}

To develop the proposed system, we use HTML5, JavaScript, and bootstrap with some related libraries. The server-side application is developed using Node.js and Express. JavaScript and JQuery are used for the dynamic processing of the pages. Recorder.js\footnote{https://github.com/mattdiamond/Recorderjs}, which is a library using Web Audio API, is used for speech recording. We use wavesurfer.js\footnote{https://wavesurfer-js.org/} as the visualization of the recorded waveform.
Bootstrap, which is a CSS framework, is used instead of PHP, as a part of the GUI, which enables web applications with dynamic processing.
The most programs in the developed system, e.g., membership registration and data transfer by form, are used in many web services, and most existing browsers support the functions.

\section{Operation of Corpus Builder}

Corpus builders first register for membership and create the account to get all the service of the system to create the target corpus and check the recording state. The required account information includes name, gender, age, place, and so on. After the registration and log-in, builders move to the corpus building page, and input corpus title and explanation as a text format. Data of the recording scripts including sentences and other information, e.g., kana and accent information in the case of Japanese, as is shown in Fig.\,\ref{fig:screen}. The data is provided in a CSV format. An initial corpus database without recorded speech is created after submitting the information.

The data format of CSV is ``name of the 1st layer, name of the 2nd layer, name of the 3rd layer, file name, sentence, accent/intonation information.'' The current system is designed only for English and Japanese, however, it is easy to extend to other languages, such as English. The layer names are used for the categorization of the speech, e.g., speaking styles and emotions. This indicates that the builder can collect speech samples of multiple categories, which is useful in the case of collecting emotional speech. To clarify the pronunciation for a difficult sentence including Kanji in Japanese, ruby can be used as a format of [kanji](ruby), e.g., [行脚](あんぎゃ). Each layer corresponds to a directory, and wave files are stored in the 3rd directory with the the respective file name.

The open pages of corpus recording for the recording participants and the closed pages to check the state and progress of the recording for the corpus builder are automatically created after the initialization of the corpus database. When a participant finished recording all the assigned sentences, builders have the option of checking the result manually or not. Followed by the optional manual check, builders send a password to the participant in the case of crowdsourcing to proceed to the payment. Finally, builders can download the full corpus data from the server after all the participants finish the recording.

\section{Operation of Recording Participant}

Participants also need to register for membership to start the recording. In the speech processing research in some topics, gender, age, accent information plays an important role. Therefore, the participants are required to input the basic personal information that doesn't identify an individual, e.g., gender, age, birthplace, living place, and so on. The information associated with each speaker can be download together with the main speech corpus. After the registration, a participant logs in to the system using the created account and access to the page specified by the corpus developer in advance. The developer also may specify the recording device, i.e., PC or smartphone, and recording environment, i.e., silent room or open place under the noisy condition.

In any device and environment, participants can conduct the recording basically with the same procedure. The current implementation assumes that the developer needs to collect speech samples in a clean condition at a low noise level. In the recording, the volume and noise levels are automatically checked and the direction of a retry is given when the conditions are not satisfied. Figure\,\ref{fig:wave} shows an example of the recorded speech waveform. When the conditions are satisfied, participants can move to the next sentence, and finish the recording by repeating the procedure. After the recording of all specified sentences, participants report the situation to the builder, and the builder checks the result. In the case of crowdsourcing, the builder sends the password for the payment and all recording procedure is finished.

\section{Conclusions}

In this paper, we briefly describe our attempt at developing a multi-platform browser-based speech recording and corpus building system to reduce the total costs for the speech processing researchers and developers. Recent advances in machine learning with computer processing bring us to increase the number and kinds of speech materials in the data-driven speech processing such as automatic speech recognition and text-to-speech synthesis. However, it is still difficult for some researchers and developers to collect a variety of speech data, e.g., speech samples with a wide range of ages, emotions, and speakers, at a low cost. The proposed system tries to alleviate such difficulties to take the form of a corpus building service at a single server. The system provides the user registration and basic functions for both corpus builder and recording participants at an open form. In the near future, the closed and the public tests will be conducted and the details of the tests and the service are presented as an article.




\end{document}